# MARVELO:
# Wireless Virtual Network Embedding for Overlay Graphs with Loops

Haitham Afifi
*Paderborn University*

Sébastien Auroux
*Paderborn University*

Holger Karl
*Paderborn University*

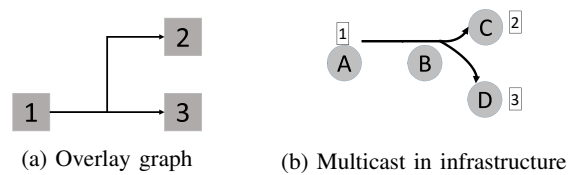

(a) Overlay graph  (b) Multicast in infrastructure

Figure 1: VNE maps processing blocks to nodes

*Abstract*—When deploying resource-intensive signal processing applications in wireless sensor or mesh networks, distributing processing blocks over multiple nodes becomes promising. Such distributed applications need to solve the placement problem (which block to run on which node), the routing problem (which link between blocks to map on which path between nodes), and the scheduling problem (which transmission is active when). We investigate a variant where the application graph may contain feedback loops and we exploit wireless networks' inherent multicast advantage. Thus, we propose Multicast-Aware Routing for Virtual network Embedding with Loops in Overlays (`MARVELO`) to find efficient solutions for scheduling and routing under a detailed interference model. We cast this as a mixed integer quadratically constrained optimisation problem and provide an efficient heuristic. Simulations show that our approach handles complex scenarios quickly.

## 1. Introduction

Wireless sensor networks (WSNs) have conventionally focussed on simple data collection applications owing to their hardware constraints. With the advent of more powerful yet still cheap hardware (e.g., Arduino or Raspberry Pis), a new class of applications for WSNs is emerging where the collected data is more voluminous and the application constraints like delays are tighter. Examples for such applications often come from the acoustic or video signal processing domain: distributed microphone arrays, collecting streams of audio data, or acoustic-based localisation of speakers. In such applications, conventional figures of merit for WSNs like energy efficiency take second place (as nodes are often wall-plugged) compared to application-oriented ones like delays, dependability, or feasibility with constrained wireless resources.

A simplistic approach to support such applications in a WSN-like network would be to record data on distributed nodes and send all that data to a central location where signal processing happens. This is, however, not a compelling solution as it introduces a single point of failure, requires substantial data rates not necessarily available over wireless links, and might result in high delays. An alternative solution could distribute individual signal processing *blocks* onto the *nodes* of the WSN and perform processing locally.

This idea of in-network processing has been considered in WSNs before but typically for much simpler applications than signal processing applications (be it acoustic or otherwise). Typically, only simple aggregation functions and low data rates were investigated. The scenario here is more challenging as the processing requirements of signal processing blocks can differ substantially, as do data rate and delay requirements between them. A similar idea of in-network processing is currently considered in the context of network function virtualisation (NFV), but this focuses on wired networks. Solutions from that field are not easily applicable due to the inherently different characteristics of wired and wireless networks.

Formally, our problem is related to *virtual network embedding* (VNE) [1]: given a wireless network modelled as an *infrastructure graph* and a distributed application modelled as an *overlay graph*, map the blocks of the overlay to nodes of the infrastructure and map links of the overlay to paths in the infrastructure, under typical node and link capacity constraints. Figure 1 illustrates an example scenario: The signal processing blocks 1, 2, and 3 are mapped to the nodes A, C, and D, respectively; block 1 sends the same data to blocks 2 and 3, which might be exploited by cleverly multicasting from B. The VNE problem has been well studied in wired networks [1]. So far, however, specific wireless properties have not been fully addresses. According to [2], we provide for the first time an exact wireless virtualisation solution that considers cyclic flows in the *overlay graph* and exploit multicast property in the *infrastructure graph*. Specifically, our contributions are:

- We support overlay graphs with loops, as often found in signal processing applications.
- We leverage the *wireless multicast advantage* as typical signal processing applications sending the same data to multiple receivers (e.g., in loops).
- When scheduling wireless transmission, we consider interference from all sending nodes.

- For this process, we provide a formal characterisation as an optimisation problem as well as heuristic solution.
- We evaluate the performance of both approaches.

We call this combination of features the **M**ulticast-**A**ware **R**outing for **V**irtual network **E**mbedding with **L**oops in **O**verlays (MARVELO) problem.

In the following Section 2, we discuss how our contributions differ from existing results. Section 3 formalises our problem and Section 4 describes the heuristic solution. Evaluation results are presented in Section 5.

## 2. Related Work

The main difference between wired and wireless VNE arise from the wireless nature when one active path interferes on neighbouring nodes [2]. This fact was, however, compensated in [3] [4] by assuming a perfect interference cancellation mechanism running at nodes, which is not necessarily available in WSNs.

The work in [5] [6] also studied the wireless VNE problem, but they assumed a limited interference model, only neighbours who directly connect to a node are considered to be interfering on this node. However, when nodes are operating in the same collision domain (i.e., same space, spectrum, time), this assumption oversimplifies the problem due to two reasons. First, the interference of one path on its neighbours may wreck the connectivity of neighbours' paths. Second, even nodes that do not have a direct connection still contribute in the interference.

Although the authors of [7] [8] considered interference from all neighbouring nodes during the placement process, their network flow model follows the flow conservation rule, ignoring the wireless multicast advantage.

In [9] the authors proposed a heuristic solution for the wireless multicast problem, while in [10] the authors proposed a MILP exact formulation for this problem. However, neither solution directly supports multi-hop flows. On the contrary, the authors of [11] formulated the multi-hop flow using a non-linear formulation. Nevertheless, they all assumed that the packet loss ratio for each wireless edge is fixed; overlooking the dependency between packet loss ratio and the signal to interference noise ratio (SINR).

## 3. The MARVELO problem

In this section, we first describe MARVELO informally. Then, we propose an exact formulation and explain how our solution can be used in applications that have feedback loops in a multicast environment.

### 3.1. Problem Definition

We define the **infrastructure graph** as $G_I = (V, T, \Gamma, C, \text{SINR}_{th}, N_o)$. Each node $v \in V$ has a capacity $c_v = C(v)$ (representing resources such as memory and CPU). Moreover, we define $v_{\text{src}}$ and $v_{\text{sink}}$ as source and sink nodes. The former can sense input signals (e.g., by a microphone); the latter could be a gateway node.

For simplicity, all nodes transmit with the same transmission power, and have an identical normalized noise floor $N_o$ (both assumptions are easy to generalize). The matrix $\Gamma$ contains the long-term average attenuation $\gamma_{v,v'}$ between any two nodes $(v, v'); v \neq v'$. We assume central knowledge of slowly varying $\Gamma$, updated within its coherence time, and consider potential fast-fading phenomenon to be handled by lower-layer mechanisms.

We assume a time-slotted model where transmissions take place in distinct time slots $t \in T$ grouped into time frames; $|T|$ is the maximum number of slots in a time frame. All nodes are perfectly synchronized to these slots. For a node to receive at time slot $t$, its SINR must exceed a given threshold $\text{SINR}_{\text{th}}$ to enable transmission at a given desired rate $R$ bit/s at negligible error rate.

**Overlay graph** denotes the distributed application as a directed graph $G_O = (P, L, W)$. Each processing block $p \in P$ requests node resources given by $w_p = W(p)$. Block $p$ sends the same data to all its successors $p'$ with $(p, p') \in L \subset P \times P$. A link $(p, p')$ needs a data rate of at most $R/|T|$ bit/s; we will schedule a link's transmission in a single time slot per time frame.[1] Similar to the infrastructure graph, we define $p_{\text{src}}$ and $p_{\text{end}}$ as the source and sink blocks of the application.

Our task is now to map blocks to nodes and overlay links to infrastructure paths. For block mapping, typical capacity constraints hold (Section 3.2.2). For link mapping, one would be tempted to use common flow conservation constraints: a node's incoming flow equals its outgoing flow unless a processing block is placed on this node.

However, this would not do justice to the goal of leveraging multicast. Let us reconsider Figure 1. Node B receives one flow from A but has to forward it to C and D. Under flow conservation rule, it could only do that if it received two flows from A, but that is wasteful (Figure 2a). Hence, we have to loosen the flow conservation restriction by allowing a node to forward *at least as much* traffic as it has received.

This relaxation, however, ensues an unfortunate consequence. Consider Figure 2b, which shows a loop of block's A traffic being forwarded among nodes D, E, and F. While conventional flow conservation rules would prevent such a loop (as block 3 on D would remove this flow), under this relaxed rule, this loop is consistent with all constraints (say, node D receives flow 1, pushes it into its block 3 but D *also* forwards it to F; flow 1 at E and F is balanced). We hence need to come up with additional constraints to prevent such loops. In doing so, we also have to deal with loops deliberately created in the overlay graph (Figure 6).

### 3.2. Optimisation problem

This section formalises our model as an optimisation problem.

---
1. Extending this model to different link rates and spreading one transmission over multiple time slots or grouping links in a time slot is not difficult but requires notation that is a bit cumbersome. As this would detract from the core points of the paper, we leave that for an extended version.

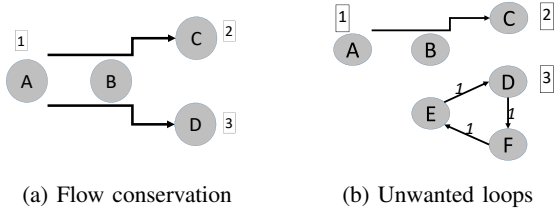

Figure 2: Wireless-VNE flow challenges

### 3.2.1. Decision variables.
Given the infrastructure and overlay graphs, we use a binary variable $\theta \in \Theta : P \times V$ for placing a processing $p$ on an actual node $v$.

Additionally, we use a binary variable $f \in F : P \times V \times T$ to indicate if the output traffic of processing block $p$ is originated or forwarded by node $v$ at time slot $t$.

We use a binary variable $s \in S : V \times V \times P \times V \times T$ which is used for scheduling and flow routing. $s(v_1, v_2, p, v_3, t) = 1$ if and only if node $v_1$ is sending to $v_2$ at time slot $t$ the output traffic of processing block $p$, which is placed on node $v_3$. Otherwise, $s()$ is equal to 0.

In fact, $s$ fully determines $\theta$ and $f$, which are mostly for conceptual and notational convenience. For additional convenience, we use a binary variable $\beta \in B : T$ to indicate if a time slot $t$ is used or not.

### 3.2.2. Constraints.
We group our constraints into four main groups. **First**, we define variable interdependency between $s$, $f$, and $\beta$ (variables $s$ and $\theta$ are coupled via the flow constraints later on). The relationship $f(p,v,t) > 0 \Leftrightarrow s(v,v_i,p,v_j,t) > 0 \,\forall p,v,t$ is expressed by the constraints (1) and (2) by means of a big-M construction. Similarly, (3) and (4) expresses $\beta(t) > 0 \Leftrightarrow \exists v_i, v_j, v_k, p : s(v_i, v_j, p, v_k, t) > 0 \,\forall t$.

$$\sum_{v_i \in V \setminus v} \sum_{v_j \in V} s(v, v_i, p, v_j, t) - f(p, v, t) \geq 0, \quad \begin{array}{l} \forall v \in V \\ \forall p \in P \\ \forall t \in T \end{array} \quad (1)$$

$$\sum_{v_i \in V \setminus v} \sum_{v_j \in V} s(v, v_i, p, v_j, t) - \mathcal{M} \cdot f(p, v, t) \leq 0, \quad \begin{array}{l} \forall v \in V \\ \forall p \in P \\ \forall t \in T \end{array} \quad (2)$$

$$\sum_{v_i \in V} \sum_{v_j \in V} \sum_{p \in P} \sum_{v_k \in V} s(v_i, v_j, p, v_k, t) - \mathcal{M} \cdot \beta(t) \leq 0 \quad \forall t \in T \quad (3)$$

$$\sum_{v_i \in V} \sum_{v_j \in V} \sum_{p \in P} \sum_{v_k \in V} s(v_i, v_j, p, v_k, t) - \beta(t) \geq 0 \quad \forall t \in T \quad (4)$$

**Second**, we ensure node mapping and adequate wireless communication in constraints (5) – (8).

In (5) and (6), we ensure that a processing block is placed only once ($p_{\text{src}}$ is placed on all source nodes and $p_{\text{sink}}$ is placed on $v_{\text{sink}}$) and the nodes' capacity constraints are not violated. Since we assume unit data rate between blocks, (7) ensures that only one block's traffic is sent by a node in one time slot. Also, a node is allowed either to transmit or receive in a given time slot (i.e., half-duplex radios). In (8), we have our only quadratic constraint, which allows transmissions from node $v$ to $v'$ if the SINR at $v'$ is bigger than or equal to $\text{SINR}_{\text{th}}$. In this check, we consider interference from all nodes except node $v$.

$$\sum_{v \in V} \theta(p, v) = 1, \quad \forall p \in P \setminus \{p_{\text{src}}\} \quad (5)$$

$$\sum_{p \in P} \theta(p, v) \cdot w_p \leq c_v, \quad \forall v \in V \quad (6)$$

$$\sum_{p \in P} f(p, v, t) + \sum_{v_i \in V} \sum_{p \in P} \sum_{v_j \in V} s(v_i, v, p, v_j, t) \leq 1, \quad \begin{array}{l} \forall v \in V \\ \forall t \in T \end{array} \quad (7)$$

$$\sum_{p \in P} \sum_{v_i \in V} s(v, v', p, v_i, t) \cdot \text{SINR}_{\text{th}} \leq \frac{\sum_{p \in P} \sum_{v_i \in V} s(v, v', p, v_i, t)}{N_o + I(v, v')},$$

$$\text{where } I(v, v') = \sum_{p \in P} \sum_{\substack{u \in V \\ u \neq v}} f(p, u, t) \cdot \gamma_{u, v'}, \quad \begin{array}{l} \forall v, v' \in V \\ \forall t \in T \end{array} \quad (8)$$

**Third** we check flow constraints in the infrastructure graph. We consider in these constraints the mapping of the overlay links $l = (p_1, p_2)$ to a flow between nodes.

Constraint (9) checks that a node $v$ has received $p_1$'s traffic, irrespective from which node $v_i$, before placing $p_2$ on node $v$. For flow control, (10) ensures that when node $v$ receives $p_1$'s traffic, it will either forward this traffic or place $p_2$ on $v$. Conversely, (11) allows node $v$ to send block $p$'s traffic only if $v$ has received $p$'s traffic or $p$ is placed on $v$. Note that when not supporting multicast, these three inequalities collapse into one equality constraint.

$$\sum_{v_i \in V} \sum_{v_j \in V} \sum_{t \in T} s(v_i, v, p_1, v_j, t) - \theta(p_2, v) \geq 0, \quad \begin{array}{l} \forall v \in V \\ \forall (p_1, p_2) \in L \end{array} \quad (9)$$

$$\sum_{v_i \in V} \sum_{t \in T} s(v_i, v, p_1, v_j, t) - \sum_{(p_1, p_2) \in L} \theta(p_2, v)$$
$$- \sum_{v_i \in V \setminus v} \sum_{t \in T} s(v, v_i, p_1, v_j, t) \leq 0, \quad \begin{array}{l} \forall v, v_j \in V \\ \forall p_1 \in P \end{array} \quad (10)$$

$$\sum_{v_i \in V} s(v, v_i, p, v', t)$$
$$- \mathcal{M} \sum_{v_i \in V \setminus v} \sum_{t_i \in T} s(v_i, v, p, v', t_i) - \mathcal{M}\theta(p, v) \leq 0, \quad \begin{array}{l} \forall v, v' \in V \\ \forall p \in P \setminus p_{\text{sink}} \\ \forall t \in T \end{array} \quad (11)$$

**Fourth**, we need to exclude loops as in Figure 2b. In that figure, e.g. $F$ cannot really send to $E$ the traffic originated by block 1 on A – because there is no active path via which $F$ could receive 1's traffic. Checking whether such a path exists is not obvious. Constraints for a maximum path length of 1 are to easy write down; it gets more and more complex the longer we allow the paths to become. Hence, lest we have to write down all these constraints manually, we use Algorithm 1 to construct these constraints systematically.

Algorithm 1's goal is to generate a constraint expressing whether a node $v_{\text{start}}$ has received block $p$'s traffic originating from node $v$. To do so, it generates all possible infrastructure paths from $v$ to $v_{\text{start}}$ via a depth-first search (DFS). For each path, a conjunctive constraint is produced to check whether *each* node on this path has received this traffic. Then, we need to check whether at least one of all these paths does carry the traffic; this is expressed by a disjunction of these conjunctions.

The challenge to expressing conjunctions and disjunctions is to find a linear form for it. A conjunction between variables $x_i, i = 1, \ldots, n$ (with 0=False and True, otherwise) can be expressed as $\frac{1}{2^n} + \sum_{i=1}^{n} \frac{x_i}{2^i} \geq 1$. A disjunction corresponds to $\frac{1}{2} + \sum_{i=1}^{n} \frac{x_i}{2} \geq 1$.

In our present case, $x_i$ corresponds to the fact that a node $v_2$ receives a particular block $p$'s traffic (placed at some $v$) from a neighbour $v_1$, irrespective of the time slot. This corresponds to $\sum_{t \in T} s(v_1, v_2, p, v, t)$ being zero or larger. These sums have to be computed, for every $p$, for $(v_1, v_2)$ along all possible paths starting from a particular node $v_\text{start}$ under consideration to $p$'s hosting node (i.e., $v$). This happens by calling Algorithm 1 recursively with parameters $(v_\text{start}, v_\text{end}, p, v, \text{visitedNodes}, r)$. visitedNodes represents nodes on the *currently* considered path from $v_\text{start}$ to $v$; $r$ is the recursively constructed conjunction along this path. Algorithm 1 constructs the conjunction sum as a continued fraction, which is simpler to do in a recursive algorithm. It produces the terms stated above. For details, please see the listing of Algorithm 1.

The result of this algorithm, called for all node combinations, is then forming the following constraint:

$$\text{trackFlow}(v_\text{start}, v_\text{end}, p, v, \{v_\text{start}, v_\text{end}\}, 1) \begin{subarray}{l} \forall v_\text{start} \in V \setminus v_\text{end} \\ \forall v_\text{end} \in V \setminus v_\text{start} \\ \forall p \in P \\ \forall v \in V \setminus \{v_\text{start}, v_\text{end}\} \\ \forall t \in T \end{subarray}$$
$$- s(v_\text{start}, v_\text{end}, p, v, t) \geq 0 \quad (12)$$

The initial value of $r$ is equal to one and any new path updates $r$ as follows $r_\text{new} = \frac{r + x_i}{2}$. Consequently, the value of $r$ falls in range $]0, 1]$.

Figure 3 depicts the algorithm's progress. Given 4 fully connected nodes, we check if node A (i.e., $v_\text{start}$) can send to node B (i.e. $v_\text{end}$) traffic of block $p$ that is originated by node C (i.e. $v$). There are two available routes from $C$ to $A$ (directly and via $D$), but only one route may be selected so that they do not conflict with (10) and (11). Each route's availability is characterised by the fractional terms on the right.

---

**Algorithm 1:** trackFlow function

**Input:** $v_1$, $v_2$, $p$, $v$, visitedNodes, $r$
**Result:** Constraint for possible paths in a DFS tree

1 **if** $v_1 = v$ **then**
2     **return** $\frac{\sum_{t \in T} s(v_1, v_2, p, v, t) + r}{2}$
3 **end**
4 visitedNodes = visitedNodes+$\{v_1\}$
5 sum = 0
6 **foreach** $v_i \in V \setminus$ visitedNodes **do**
7     $r_\text{new} = \frac{\sum_{t \in T} s(v_i, v_1, p, v, t) + r}{2}$
8     sum = sum
       + trackFlow $(v_i, v_1, p, v, \text{visitedNodes}, r_\text{new})$
9 **end**
10 **return** sum

---

### 3.3. Objective

Our **objective** is to minimize the number of used time slots; $\min \sum_{t \in T} \beta(t)$. This reflects latency requirements of typical signal-processing or real-time applications.

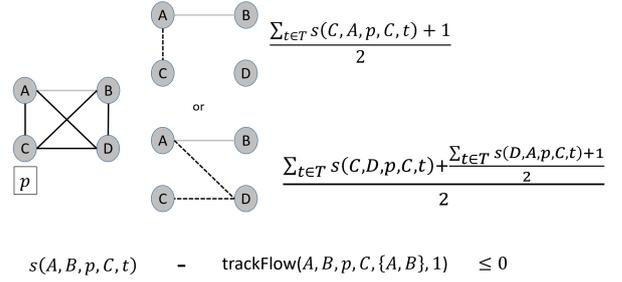

Figure 3: Example for Algorithm 1: $p$ is placed on $C$, can $A$ send to $B$ the traffic of $p$ originated by $C$?

## 4. A heuristic for MARVELO

VNE problems are NP [1] and finding an optimal solution is too slow. Therefore, we develop a heuristic.

### 4.1. Heuristic overview

The core idea is to start from the source block, progress from overlay link to link in a topological order, mapping the link to a path in the infrastructure and mapping blocks to nodes in the same step. More precisely, when we map a link $(p_1, p_2)$, there are two cases: (1) The receiving block has already been placed, then link mapping just means finding a path between the two nodes hosting $p_1$ and $p_2$. (2) If $p_2$ is not yet placed, we also have to find a hosting node for $p_2$, jointly with finding a path towards that node.

We can hence think about this as a link mapping problem, where we progress from link to link. Whenever a link has been mapped, we have choices for mapping the next link (to different paths, or to different nodes and paths). This is a search problem in a tree of possible link mapping decisions. A sequence of link mapping decisions that maps all links constitutes a feasible solution.

A brute-force algorithm to find the optimal solution would have to explore this tree in its entirety. This is clearly not feasible. Hence, we introduce three control mechanisms to limit the search space: lookahead, backtracking, and degree of this search tree.

### 4.2. Lookahead level

Suppose we have committed to the mapping of a link $l_1$. To determine how to map the next link $l_2$, we could just look at the options for this link and pick the best possible option. In addition, we could also *look ahead*: We consider all possible options for mapping the next *level* many links, exploring an entire subtree. Among those possible mapping combinations, we choose the best one and map *level* many links in a single step. Figure 4 shows examples for 1-level and 2-level lookahead mapping (note that numbers in these figures indicate the total number of used time slots and the circles are *link mappings*, not nodes!). When *level* equals the number of links, this scheme degenerates into exhaustive brute-force search.

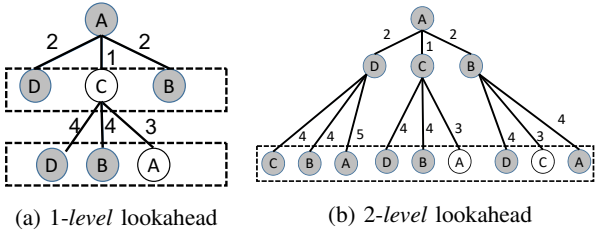

(a) 1-*level* lookahead   (b) 2-*level* lookahead

Figure 4: Looking ahead in the link mapping search tree (dashed boxes indicate levels of decision making)

### 4.3. Backtracking

Suppose we have mapped *level* many links and try to find a solution for the next *level* links. What happens if no feasible solution can be found? As is typically done, we backtrack, reject the decisions taken for the previous mapping of *level* many links, and start again with the remaining best solution. This is illustrated in Figure 5.

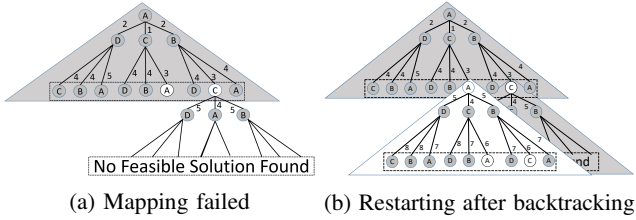

(a) Mapping failed   (b) Restarting after backtracking

Figure 5: Backtracking in the link mapping search tree

### 4.4. Limit tree degree

As a second parameter, we limit the degree of the link mapping search tree. When mapping a link $(p_1, p_2)$ with $p_1$ hosted on $v_1$, and trying to find a node $v_2$ to host $p_2$, we only look at the $k$ neighbors of $v_1$ with best attenuation values.

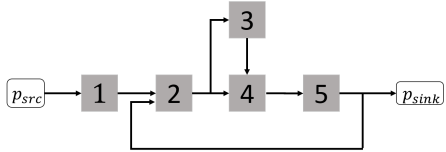

Figure 6: Overlay graph from acoustic signal processing [12] with a loop from block 5 to 2

## 5. Evaluation

### 5.1. Scenario

The exact formulation is solved using Gurobi Optimizer 7.5; our heuristic is implemented in Python. All simulations were executed in single-threaded mode on Intel Xeon X560 cores running at 2.67 GHz.

Our simulated environment consists of a room with area $25 \times 25$ m$^2$, where nodes are placed uniformly at random, independently from each other. The attenuation between two nodes $v, v' \in V$ is given by $\gamma_{v,v'} = \frac{1}{d_{v,v'}^2}$, where $d_{v,v'}$ is the distance between the two nodes.

We vary the number of nodes and run 50 independent realizations for each number of nodes. In each realization, node capacities are uniformly distributed $c(v) \sim U(\max(w_p), \sum_{p \in P} w_p)$. Furthermore, $v_{\text{src}}$ and $v_{\text{sink}}$ nodes are picked randomly per realization.

We choose a generic algorithm from the field of acoustic signal processing for our overlay graph [12]. Figure 6 depicts the graph with 5 processing blocks equally weighted $w_p = w_o$. We add two artificial blocks $p_{\text{src}}$ and $p_{\text{sink}}$ to assign to $v_{\text{src}}$ and $v_{\text{sink}}$; $w_{p_{\text{src}}} = w_{p_{\text{sink}}} = 0$.

### 5.2. Execution Time

Figures 7 and 8 evaluate our heuristic's median runtime for different configuration setups. First, in Figure 7 we set the lookahead *level* to 1 and vary the search tree degree $k$ between 3, 6, and all neighbours. We observe that the runtime increases exponentially as the number of nodes increases. Moreover, limiting the search space to $k$ neighbours reduces the execution time significantly as $k$ decreases.

We investigate the impact of increasing the lookahead *level* in Figure 8. We observe that increasing the *level* from 1 to 2 has a higher impact than limiting the neighbourhood to $k$. The optimal (zero gap) solution's median execution time over 50 runs is 140 seconds for 4 nodes and 2564 seconds for 6 nodes. Therefore, we limit the exact model's evaluation to 6 nodes.

### 5.3. Schedule length: Heuristic vs. optimal solution

We compare the heuristic and the exact model using the heuristic gap $\frac{\sum_{t \in T} \beta_h(t) - \sum_{t \in T} \beta_{opt}(t)}{\sum_{t \in T} \beta_{opt}(t)}$, where $\beta_h$ and $\beta_{\text{opt}}$ are the heuristic and optimal (with no gap) solution's schedule length.

In Figures 9 and 10, we show the 95% confidence interval for the mean gap with 4 and 6 nodes for different $k$. When using 1-*level* lookahead (in Figure 9) or 2-*level* (in Figure 10), limiting the search space to the nearest 3 neighbours improves the optimality gap.

The reason is random selection during link mapping, when two or more mapping give the same minimal additional number of time slots. As the search space shrinks (i.e., decreasing $k$), such randomness becomes more guided towards nearing nodes.

For the same number of nodes and $k$, the 1-*level* lookahead has lower gap than 2-*level*. Hence, increasing the *level* number does not necessarily yield lower number of time slots (except when $level = |L| \leftrightarrow$ brute-force).

### 5.4. Schedule length in large scenarios

We further investigate the schedule length using the heuristic for many nodes. Figures 11 and 12 show that (a) the average schedule length does not change significantly with the number of nodes, (b) the *level* number has an impact; 1-*level* has smaller mean gap than 2-*level* lookahead. This confirms our hypothesis in Section 5.3: increasing the *level* number does not always yield a better solution.

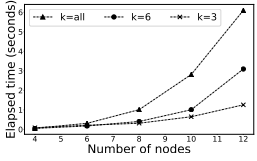

Figure 7: Execution time for k-neighbours

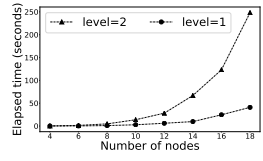

Figure 8: n-*level* execution time

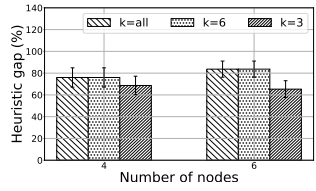

Figure 9: 1-*level* gap

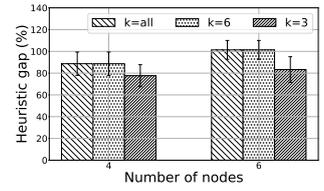

Figure 10: 2-*level* gap

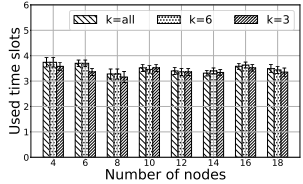

Figure 11: 1-*level* used slots

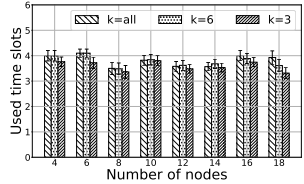

Figure 12: 2-*level* used slots

Furthermore, we observe that limiting $k$ yields (in most cases) a shorter schedule. An exception would be in Figure 11, $k = 6$ and 14 nodes yields longer schedules than considering all neighbouring nodes.

The number of utilized nodes (for hosting blocks or forwarding) interestingly has no clear dependence on either level, $k$, or node number (results skipped due to page limitations).

## 6. Conclusion

We have introduced a new formulation for the wireless VNE problem, suitable for multicast multi-hop environments. Since optimally solving such a problem is time consuming, we also proposed a heuristic algorithm that can be controlled using two parameters; *level* and k nearest neighbours. We have shown that changing both parameters can have a significant impact on the execution time, especially for the *level* parameter.

Although setting the *level* parameter to the number of links in the overlay network introduce an optimal solution, decreasing *level* does not always yield a worse solution. We have shown that the 1-*level* setup is having a better lower mean of used time slots than 2-*level*. On the other hand, reducing the search space to the nearest k-neighbours does not have a substantial impact on the number of used time slots.

A typical use case of our analysis is for applications with fast embedding requirements. We have shown that our heuristic can get acceptable results using a setup that requires low execution time. In this case, a quick solution can be found for the wireless VNE problem. Then, it can be optimized by changing the *level* and k-neighbours parameters if more time is provided.

## Acknowledgments

The authors gratefully acknowledge support from the DFG research group Acoustic Sensor Networks FOR 2457.

## References

[1] A. Fischer, J. F. Botero, M. T. Beck, H. de Meer, and X. Hesselbach. Virtual network embedding: A survey. *IEEE Communications Surveys Tutorials*, 15(4):1888–1906, 2013.

[2] J. van de Belt, H. Ahmadi, and L. E. Doyle. Defining and surveying wireless link virtualization and wireless network virtualization. *IEEE Communications Surveys Tutorials*, 19(3):1603–1627, 2017.

[3] J. van de Belt, H. Ahmadi, and L. Doyle. A dynamic embedding algorithm for wireless network virtualization. 06 2014.

[4] M. Selimi, L. Cerd-Alabern, M. Snchez-Artigas, F. Freitag, and L. Veiga. Practical service placement approach for microservices architecture. In *2017 17th IEEE/ACM International Symposium on Cluster, Cloud and Grid Computing (CCGRID)*, pages 401–410, May 2017.

[5] S. Abdelwahab, B. Hamdaoui, M. Guizani, and T. Znati. Efficient virtual network embedding with backtrack avoidance for dynamic wireless networks. *IEEE Transactions on Wireless Communications*, 15(4):2669–2683, April 2016.

[6] R. Riggio, A. Bradai, D. Harutyunyan, T. Rasheed, and T. Ahmed. Scheduling wireless virtual networks functions. *IEEE Transactions on Network and Service Management*, 13(2):240–252, June 2016.

[7] G. Di Stasi, S. Avallone, and R. Canonico. Virtual network embedding in wireless mesh networks through reconfiguration of channels. In *IEEE 9th Intl. Conf. on Wireless and Mobile Computing, Networking and Communications (WiMob)*, pages 537–544, Oct 2013.

[8] M. Li, X. Wang, S. Chen, M. Song, and Y. Ma. *Virtual Mapping of Software Defined Wireless Network Based on Topological Perception*, pages 238–249. Springer International Publishing, 2016.

[9] P. Lv, Z. Cai, J. Xu, and M. Xu. Multicast service-oriented virtual network embedding in wireless mesh networks. *IEEE Communications Letters*, 16(3):375–377, March 2012.

[10] X. Gao, W. Zhong, Z. Ye, Y. Zhao, J. Fan, H. Cao, H. Yu, and C. Qiao. Virtual network mapping for reliable multicast services with max-min fairness. In *2015 IEEE Global Communications Conference (GLOBECOM)*, pages 1–6, Dec 2015.

[11] M. Li, C. Hua, C. Chen, and X. Guan. Application-driven virtual network embedding for industrial wireless sensor networks. In *IEEE Intl. Conf. on Communications (ICC)*, pages 1–6, May 2017.

[12] J. Schmalenstroeer, J. Heymann, L. Drude, C. Boeddecker, and R. Haeb-Umbach. Multi-stage coherence drift based sampling rate synchronization for acoustic beamforming. *IEEE 19th International Workshop on Multimedia Signal Processing (MMSP)*, 2017.